\documentclass[11pt,amsart]{article}
\usepackage{bm}
\usepackage{graphicx}
\usepackage{graphics}
\usepackage{float}
\usepackage{amsmath}
\usepackage{amssymb}
\usepackage{amsfonts}
\usepackage{latexsym}
\usepackage{epsfig}
\usepackage{dcolumn}
\usepackage{url}

\usepackage[titles]{tocloft} 

\usepackage[francais,english]{babel}
\usepackage{lmodern} 
\usepackage[T1]{fontenc}
\usepackage{pdfsync}
\usepackage[pdftex,bookmarks,colorlinks,breaklinks]{hyperref}
\hypersetup{linkcolor=black,
citecolor=blue,filecolor=red}

\newcommand{\Ecal}{{\mathcal E}}

\newcommand{\Rcal}{{\mathcal R}}

\newcommand{\Tcal}{{\mathcal T}}

\newcommand{\Fcal}{{\mathcal F}}

\newcommand{\bvec}{{\textbf b}}

\newcommand{\mvec}{{\textbf m}}

\newcommand{\kvec}{{\textbf k}}
\newcommand{\pvec}{{\textbf p}}

\newcommand{\muvec}{{\boldsymbol \mu}}

\usepackage{epstopdf}

\title{\sf defects in quasicrystals, revisited  \\ II$-$ perfect and imperfect dislocations}

\author{\textsf{Maurice Kleman}\footnote{\textsf{kleman@ipgp.fr}} 
 \vspace {15pt}\\
\small Institut de Physique du Globe de Paris $-$ 
 Sorbonne Paris Cité\\
 \small 1, rue Jussieu - Paris cedex 05, France}
  
\begin{document}
\maketitle

\scriptsize
\tableofcontents
\newpage
\normalsize

\begin{abstract}
 In this paper, the second part of a survey of the geometric properties of defects in quasicrystals studied from the Volterra viewpoint (see ref. \cite{kleman13b}), we show that: 1$- $ a {\sf disvection line} L$_{||} \subset \mathrm E_{||}$ of Burgers vector $\bvec =\bvec_{||}+\bvec_\bot $ splits naturally along  L$_{||}$ into a {\sf perfect dislocation} of Burgers vector $\bvec_{||}$ and an {\sf imperfect dislocation} of Burgers vector related to $\bvec_{\bot}$, akin to a stacking fault, (a 'phason' defect), 2$- $ the 'phason' defects are classified according to the relative position of $\Sigma_{\bot}$ with respect to a partition of the acceptance window AW which depends on the direction of $\bvec_\bot $. The perpendicular cut surface $\Sigma_{\bot}\subset \mathrm {AW}$ here introduced is a mapping of the usual cut surface $\Sigma_{||}\subset\mathrm E_{||}$. Imperfect dislocations in QCs are somewhat similar to Kronberg's synchroshear dislocations. It is also shown that climb must generically be easier than glide.
\end{abstract}\bigskip


\section{Introduction} \label{int}

It has been shown in a previous paper \cite{kleman13b} (hereunder denoted (I)) that quasicrystal rational approximants result from a periodic distribution of flips on the parent QC structure. When only one flip per unit cell, one gets the so-called Fibonacci approximants. A flip is not a topological defect and can be split into two opposite matching faults, which are topological defects; this is most probably the situation that occurs in stable approximants. A matching fault has all the characteristics of a usual stacking fault in a crystal. As we shall see, the line bordering the fault carries an effective Burgers vector $\bvec_{||}^{eff}\subset\mathrm E_{||}$ which expression derives from a geometrical analysis in E$_\bot$, and the fault displacement itself amounts to a flip.\footnote{The notations are the same as in (I).}  

Our approach to the dislocation geometrical conformations consists in 1- a Volterra process (VP) \cite{friedel64} in E$_{||}$ that yields \textsf{perfect dislocation} components, 2- an operation in E$_{\bot}$ that yields \textsf{imperfect dislocation} components. This is in contrast with the usual approach, where the defect resulting from a $\bvec= \bvec_{||}+\bvec_\bot$ dislocation is analyzed with a unique VP in $\Ecal=\mathrm E_{||}\times\mathrm E_{\bot}$; the defect observed in the physical space E$_{||}$ is then the intersection of the defect in $\Ecal$ and E$_{||}$.  
But this method, employed e.g. in \cite{kleman91b} and made easy to-day by the powerful computer simulations at our disposal, does not demonstrate at once why a {\sf disvection} (as we call the set of defects carried by a dislocation $\mathrm L_{||}\subset \mathrm E_{||}$, Burgers vector $\bvec$, see Sect.~\ref{discu}) is split in E$_{||}$ into two types of line defects, akin respectively to {perfect dislocations} (all of Burgers vectors $\bvec_{||}$) and {imperfect dislocations} (whose Burgers vectors measured in E$_{||}$ depend on $\bvec_\bot$, but vary in a subtle way according to the position of the line defect).  \smallskip

Sect.~\ref{tfs} and \ref{tfp} introduce the main concept at work in this paper, namely the distinction between {\sf true sites} and {\sf false sites} of the cut surface $\Sigma_{||}\subset \mathrm E_{||}$ for a defect line L$_{||}$; their consideration leads straight to the distinction between perfect and imperfect dislocations (matching faults). Perfect and imperfect dislocations do not mix along L$_{||}$; they separate in space. This is indeed what is observed empirically. 

Sect.~\ref{fl} and \ref{flg} develop a geometric method to obtain the Burgers vector of a matching fault; it depends on $\bvec_\bot$ and on the location of L$_{||}$. The essential ingredient is the concept of {\sf flipping vector}.

In Sect.~\ref{gc} we show that, within the present analysis, climb appears as easier than glide, a result well attested experimentally \cite{caillard00}.

Some of the results here presented were already developed in \cite{kleman03a}, from which a few figures are adapted.

 \section{Perfect and imperfect dislocations in a QC} \label{rel}
\subsection{True sites and false sites in $\mathrm E_{||}$} \label{tfs}

A VP performed in a 3D {\it periodic} crystal moves any atom $\{m\}$ ocupying any site $\mvec \in \Sigma$, the cut surface, to a site $\muvec =\mvec + \bvec$ either occupied by an atom $\{\mu\}$ crystallographically equivalent to $\{m\}$, $-$ this is a {\sf perfect dislocation} $-$, or occupied by an unequivalent atom (or not occupied at all), $-$ this is an {\sf imperfect dislocation}. We call $\mvec$ a {\sf true site} in the first case, a {\sf false site} otherwise; all the sites are either false or true, depending on $\bvec$. 
\begin{figure}[h] 
   \centering
   \includegraphics[width=3.5in]{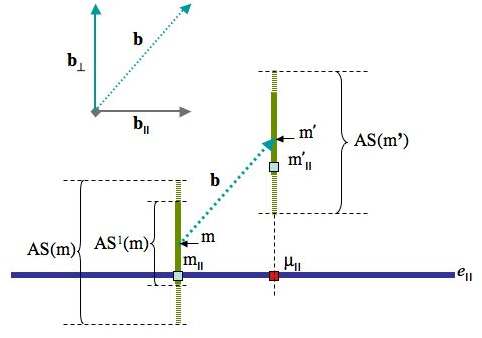} 
   \caption{Schematic representation in $P(\mvec_{||})$ of the VP displacement that affects an atom $\{m\}$ occupying the site $\mvec_{||} \in \mathrm E_{||}$
on the cut surface $\Sigma_{||}$. The sketch is made in the 2D plane $P(\mvec_{||} $), defined in the upper left corner by the directions along $\bvec_{||}$
 and $\bvec_\bot$.  
 The atom $\{m\}$ formerly in $\mvec_{||} $
 hits a site $\muvec_{||}  = \mvec_{||}  + \bvec_{||} $ which is empty in the present figure. AS(m') =AS(m)+ $\bvec$ does not intersect E$_{||}$; $\mvec_{||}$ is a false site.}
   \label{fig1}
\end{figure}

The same distinction can be made in a quasicrystal for the physical Burgers vector $\bvec_{||}$, but in this case the same cut surface $\Sigma_{||}$ can accommodate both types of sites. Figure~\ref{fig1} represents the case of a false site; let $P(\mvec_{||})\subset \Ecal$ a 2D plane that contains the QC site $\mvec_{||}$ and the two directions along $\bvec_{||}$ and $\bvec_{\bot}$; they play different roles and are thus conveniently separated. AS(m) is the atomic surface for the atom $\{m\}$ located at the site $\mvec_{||}$, $\mvec$ is the center of the hyperlattice cell to which AS(m) is attached. AS(m) intersects $P(\mvec_{||})$ along a segment denoted AS$^1$(m), smaller in length than or equal to the span of AS(m) projected onto $P(\mvec_{||})$, AS$^1$(m) $\in$ AS(m). $\mvec' = \mvec +\bvec$; in Fig.~\ref{fig1} AS(m') does not intersect E$_{||}$, $\mvec_{||}$ is thus a false site.\footnote{ In Fig. 1 {$e_{||}$ is the intersection of E$_{||}$ and $P(\mvec_{||})$; it has one intersection at most with any AS(m). Since $P(\mvec_{||})$ can be generated by a set of lines parallel to $e_{||}$, its intersection with AS(m) is a segment of curve, in fact a segment of line parallel to $\bvec_\bot$, since $\bvec_\bot \subset$ AS(m). These geometrical properties do not depend on the dimensionality of AS(m).}}

\subsection{True sites and false sites mapped inside the acceptance window $\mathrm {AW}$} \label{tfp}

\begin{figure}[h] 
   \centering
   \includegraphics[width=3.2in]{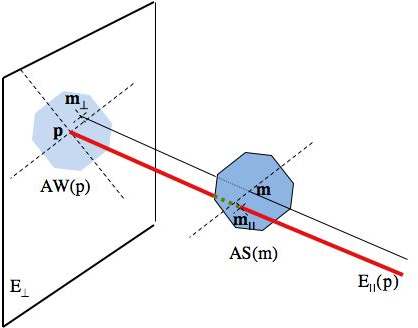} 
   \caption{Octagonal symmetry, d = 4. The 2-dimensional physical plane E$_{||}(\pvec)$, that
intersects orthogonally E$_\bot$ in one point only, $\pvec$, is represented as a line. The acceptance window AW(\pvec) is the closure of the hypercube center projections $\mvec_\bot$ whose
attached AS(m)s intersect E$_{||}$. These projections fill AW densely.}
   \label{fig2}
\end{figure}

Now we characterize the true and false sites by their images $\mvec_\bot$ into the acceptance window AW$\subset\mathrm E_\bot$. For the sake of illustration we assume in the sequel that the hyperspace is 4-dimensional, $d=4,\ d_{||}= d_{\bot}=2$, and that the symmetry is octagonal; the results extend easily to any dimension $d = d_{||}+d_{\bot}$. Fig.~\ref{fig2} indicates how a site $\mvec_{||} \in\mathrm E_{||}$ maps on $\mvec_\bot\in \mathrm {AW}$: a lift $\mvec_{||} \rightarrow \mvec$ in the same atomic surface AS(m), followed by an orthogonal projection $\mvec \rightarrow \mvec_\bot$.

Let $\bvec_{||}$ be the Burgers vector of a loop L$_{||} \subset \mathrm E_{||}$ ($ \dim{\mathrm L_{||}} =d_{||}-2$), $\bvec_{\bot}$ the corresponding unique perpendicular component. Consider all the sites $\mvec_{||}$ attached to the cut surface $\Sigma_{||}$ ($\dim{\Sigma_{||}}=d_{||}-1$). The full cut surface in the $d$-dimensional hyperspace E is $\Sigma= \Sigma_{||}\times\mathrm E_\bot$, where a copy of $\mathrm E_\bot$ is attached to each point of $\Sigma_{||}$ \cite{kleman91b}. Each copy carries a site $\mvec$, which is a vertex of the hyperlattice if it contains such a site, and if not is a point of the cell defined by interpolation from the set of nearest vertices belonging to $\Sigma$. Thus these sites $\mvec$ constitute a subset S of $\Sigma$, $\dim{\mathrm S}=d_{||} -1$, which appears as a lift of $\Sigma_{||}$ in  $\Sigma$. S projects into E$_\bot$ as a continuous domain $\Sigma_{\bot}\subset$ AW, $\dim{\Sigma_{\bot}} = d_{||}-1$, whose sites can be qualified of true or false whether they derive from a true or false site $\mvec_{||}$. The continuity of $\Sigma_{\bot}$ is proved in Appendix A; also, by its construction which confines it inside the finite manifold AW, $\Sigma_{\bot}$ might oscillate a large number of times (scaling as the number of cells traversed by $\Sigma_{||}$); furthermore the same point in AW is a possible projection of several sites $\mvec$ belonging to $\Sigma$ and defined by the interpolation alluded to above and in the Appendix A. These complications do not invalidate the reasonings that follow.
Denote $\partial \Sigma_\bot =$ L$_\bot$ the boundary of $ \Sigma_\bot$; even though $\Sigma_\bot$ can be chosen at will, its border L$_\bot$, which is the image of L$_{||}$, is fixed. 

\begin{figure}[h] 
   \centering
   \includegraphics[width=2.6in]{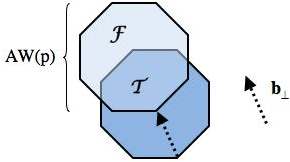} 
   \caption{For the $\bvec_\bot$ vector indicated, the acceptance window divides into two regions: $\Tcal$, where the  sites are true, and $\Fcal$, where the sites are false. See text.}
   \label{fig3}
\end{figure}

The true sites belong to a subset of AW, denoted $\Tcal$, and the false sites to its complement $\Fcal$ in AW, see Fig.~\ref{fig3}.\footnote{This figure rectifies an error in fig. 7, ref. \cite{kleman03a}.} $\Tcal$ is the intersection of AW and of a copy of AW translated by $-\bvec_\bot$. It is easy to check that, if it is so, any point $\mvec_\bot \in \Tcal$ is displaced by the VP to a point $\mvec'_\bot =\mvec_\bot+ \bvec_\bot$ which is still in AW; it is a true site since the attached atomic surface AS(m') intersects E$_{||}$. Likewise any point $\mvec_\bot \in \Fcal$ is displaced to a point outside AW and is a false site.
 
\begin{figure}[h] 
   \centering
   \includegraphics[width=3.2in]{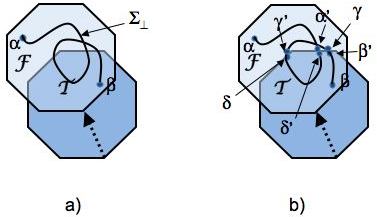} 
   \caption{a) dipole with two opposite pointlike dislocations $\bm{\alpha,\beta}$, cut surface $\Sigma_\bot$, here a line segment; b) the same dipole split into two perfect dipoles: $\bm{\beta\beta '}$ and $\bm{\delta\delta '}$ and two matching faults: $\bm{\alpha\alpha '}$ and $\bm{\gamma\gamma '}$.}
   \label{fig4}
\end{figure}

 If $\Sigma_{\bot}\subset \mathrm{AW}$ is entirely in $\Tcal$ (resp. $\Fcal$), then L$_\bot$ is entirely in $\Tcal$ (resp. $\Fcal$), i.e. L$_\bot$ is true, there are no matching faults attending the perfect dislocation $\bvec_{||}$  (resp. L$_\bot$ is false and the dislocation is imperfect). Generically a disvection can be separated into perfect and imperfect dislocations; this is sketched Fig.~\ref{fig4} for the octagonal case, with pointlike dislocations: the intersections of $\Sigma_{\bot}$ with the boundary $\{\Tcal,\Fcal\}$ between $\Tcal$
 and $\Fcal$ are replaced by two point dislocations of opposite signs, such that the successive segments $\bm{\beta\beta '}$, $\bm{\gamma \gamma '}$, $\bm{\delta\delta'}$, $\bm{\alpha'\alpha}$ terminate on pairs of point dislocations, that form dipoles. In this figure $\Sigma_{\bot}$ oscillates only once, and there is only one self-intersection $-$ this is enough to make visible that, in spite of multiple projections, the topology of $\Sigma_{\bot}$ and its self-intersections are sufficient to partition $\Sigma_{\bot}$ into perfect and imperfect sub$\Sigma_{\bot}$s, i.e.into perfect and imperfect dislocation dipoles. The only condition necessary to perform such a partition is that $\Sigma_{\bot}$ be continuous.
 
 This construction extends easily to a 3-dimensional acceptance domain; the generalization consists in
introducing two dislocation segments of opposite signs along the boundary $\{\Tcal,\Fcal\}$, where it is crossed along line segments by $\Sigma_{\bot}$. 

\subsection{False sites and matching faults: flipping vectors}\label{fl}

\begin{figure}[h] 
   \centering
   \includegraphics[width=4.25in]{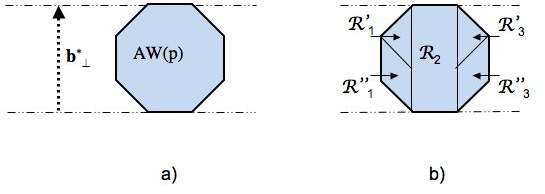} 
   \caption{Octagonal case. a) one of the four basic flipping vectors in P$_\bot$, b) partition of AW related to $\bvec_\bot^*$, see text.}
   \label{fig5}
\end{figure}

Figure \ref{fig3}
shows that if $\bvec_{\bot}$ is equal to the span of AW(p) or larger, then $\Tcal = \emptyset$, AW $=\Fcal$; the disvection is reduced to a matching fault, whatever the shape of L$_{||}$ may be. 
Let us define more precisely the relation between such a $\bvec_{\bot}$ $-$ we take it equal to the span $\bvec_{\bot}^*$ in Fig.~\ref{fig5}a $-$ and an imperfect dislocation. 

In the octagonal case there are four such vectors; they will be denoted (for later use) as in Fig.~\ref{fig9}. The corresponding hypervectors $\bvec^*=\bvec_{||} ^*+\bvec_\bot ^*$ join parallel edges of the 4D hypercells; in the general case of a $d$-hyperspace they join opposite ($d_\bot - 1$)-dimensional faces.  

\noindent They can be called {\sf flipping vectors}, since they put in correspondence two points on the boundary of AW(p), one of which deriving from a false $\mvec_{||}$ for the Burgers vector $\bvec^*$, the other one from an empty site $\muvec_{||} = \mvec_{||} + \bvec^*_{||}$. Notice that in the Fig.~2 of (I) the vector joining the centers of the atomic surfaces $\mathrm{C^-_3}$ and $\mathrm{C^+_3}$ is such a vector. 

A flipping vector $\bvec^*$ is generically a vector that joins the centers of two hypercubes that have a ($d_\bot - 1$)-dimensional face in common; they are tabulated in \cite{frenkel86} for the Penrose and the icosahedral QCs.

\begin{figure}[h] 
   \centering
   \includegraphics[width=4.7in]{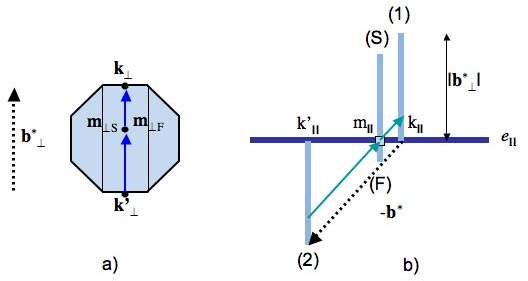} 
   \caption{Octagonal case. a) displacement of $\{m\}$ in E$_\bot$ from $\mvec_{\bot S}$ to $\mvec_{\bot F}$; here $\mvec_{\bot S}$ and $\mvec_{\bot F}$ occupy the same site $\mvec_{\bot }$;  b) displacement of $\{m\}$ in $P(\mvec_{||})$, see text.}
   \label{fig6}
\end{figure} 
What is then a \textsf{matching fault whose Burgers vector is a flipping vector}? We are guided in this search by the two representations we have already used for the VP displacement of an atomic surface AS(m) with $\{m\} \in \Sigma$, one in the perpendicular space E$_\bot$ $-$ the displacement of $\mvec_{\bot}$ $-$, the other in $P(\mvec_{||})$ $-$ the displacement of $\mvec_{||}$, both rather simple in the case in view. The $\mvec_{\bot}$ representation will appear more manageable in the general case.

There are several situations according to the position of $\mvec_{\bot}$ in AW, see the relevant partition in Fig.~\ref{fig5}b. Here we restrict to the case when the site on the cut surface belongs to $\Rcal_2$. The other cases, somehow more complicated, are discussed in Appendix B.

Consider therefore an atom $\{m\}$ represented in AW by the site $\mvec_{\bot S} \in \Rcal_2$, which suffers the VP displacement $\bvec_\bot^*$, Fig.~\ref{fig6}a. This displacement brings 
$\{m\}$ to a site $\kvec_\bot$ on the boundary of AW, where it has to flip to the site $\kvec'_\bot =\kvec_\bot - \bvec_\bot^*$; it then continues its displacement up to a final site $\mvec_{\bot F}$ to complete it to the value of the Burgers vector, the flipping displacement $\kvec_\bot\kvec'_\bot$ not being taken into account: in the present case the final site $\mvec_{\bot F}=\mvec_{\bot S}$, the total {\sf shift} in E$_\bot$ is thus equal to the flipping dispacement.

The same result holds when one considers the displacement in $P(\mvec_{||})$, see Fig.~\ref{fig6}b: AS(m) is represented by its intersection with $P(\mvec_{||})$, whose span is exactly $|\bvec^*_\bot |$; it is displaced by the VP from position (S) where AS(m) intersects E$_{||}$ in $m_{||}$ to position (1) from which it flips to (2) and then completes its displacement by going to (F), which is also (S) in the present case. The total displacement, not taking the flip into account, vanishes. 

In conclusion the Burgers vector of the dislocation in physical space is $\bvec^{*}_{||}$, and is attended by a matching fault whose shift is $\bvec^{*}_{||}$. This is much comparable to a partial dislocation in a periodic crystal; this is also the simplest case of imperfect dislocation one can meet in a QC.

\subsection{VP for general matching faults} \label{flg}

\begin{figure}[h] 
   \centering
\includegraphics[width=1.75in]{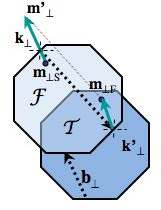} 
   \caption{VP for a false site $\mvec_{\bot S}$. The Burgers vector $\bvec_{\bot}$, which would transport $\mvec_{\bot S}$ to $\mvec_{\bot}'$, outside AW, is splitted into two parts, $\mvec_{\bot S}\kvec_\bot$ and $\kvec_\bot ' \mvec_{\bot F}$; $\kvec_\bot$ and $\kvec_\bot '$ are on the boundary of AW and related by the flip $\kvec_\bot\kvec_\bot '$.}
   \label{fig7}
\end{figure}

The general case for $|\bvec_\bot|<|\bvec_\bot^*|$ can be treated on the same basis as when $\bvec_\bot$ is a flipping vector. Starting from the site $\mvec_{\bot S} \in \Fcal$, $\bvec_\bot$ hits the boundary in $\kvec_\bot $, flips to $\kvec'_\bot$, from which site it reaches $\mvec_{\bot F}$, see Fig.~\ref{fig7}. Notice that $\mvec_{\bot F}$ is necessarily inside AW, since by construction the length spanned along the direction $\kvec'_\bot\mvec_{\bot F}$ inside AW is larger than $|\bvec\bot|$. 

Thus, in the case of the octagonal AW of Fig.~\ref{fig7}:
\begin{equation}
\label{e1}
\bvec_\bot^{eff} = \bvec_\bot -\bvec_\bot^1 \quad \rightarrow \quad \bvec_{||}^{eff} = \bvec_{||} -\bvec_{||}^1.
\end{equation}

\begin{figure}[h] 
   \centering
\includegraphics[width=3in]{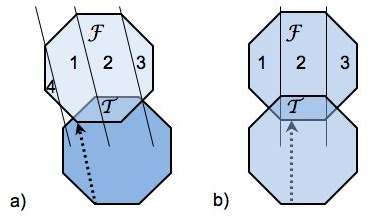} 
   \caption{The various types of matching faults: a) for a Burgers vector $\bvec_\bot$ oblique with respect to all the edges, b) for a Burgers vector parallel to one of the edges.}
   \label{fig8}
\end{figure}

The position of $\mvec_{\bot S}$ in $\Fcal$ determines which edge of the octagon $\bvec_\bot$ hits: there are four types of matching faults associated to $\bvec_\bot$, whose Burgers vectors are, \begin{equation}
\label{e2}
\bvec_{||} -\bvec_{||}^1,\bvec_{||} -\bvec_{||}^2,\ \bvec_{||} -\bvec_{||}^3, \ \bvec_{||} +\bvec_{||}^4,
\end{equation} (cf. Fig.~\ref{fig9} for the orientations of the flipping vectors), and whose corresponding fault shifts are $ -\bvec_{||}^1,\  -\bvec_{||}^2, \ -\bvec_{||}^3, \ +\bvec_{||}^4$. The corresponding domains are sketched Fig.\ref{fig8}a.
 If $\bvec_\bot$ is parallel to one of the edges,  there are only three exits possible for $\bvec_\bot$, Fig.~\ref{fig8}b. These results should apply equally well to approximants, i.e. to their metadislocations; see indeed  in \cite{gratias12} the discussion of a case similar to Fig.~\ref{fig8} (with an approach specific to an approximant) yielding a quite similar result.
 
 \section{Glide and climb} \label{gc}
 \subsection{Glide}

 In accordance with the standard definition of the glide plane of a
dislocation line in a classic 3D crystal $-$ the glide plane contains the
dislocation line and the direction of the Burgers vector $-$ we define the
glide manifold $\mathrm G= \mathrm L\times< \bvec>$ of a dislocation in a $d$-dimensional crystal as the
cartesian product of the $(d-2)$-dimensional dislocation hyperline L = L$_{||}\times \mathrm E _\bot$ (cf. \cite{kleman91b}) by the straight line along the direction of the 
Burgers vector $\bvec$.\footnote{<{\bf a}> denotes a unsigned infinite line in the
direction of {\bf a}.} This definition
holds for a straight dislocation or along a small segment of a curved dislocation.

Any movement of L in G along $\bvec$ 
 is accompanied by a movement of L$_{||}$ in G$_{||} =
\mathrm L_{||} \times <\bvec_{||}>$, the glide plane in E$_{||}$, along $\bvec_{||}$ (more precisely along the direction of the edge component $\bvec_e$ of $\bvec_{||}$). Indeed, the glide manifold G in E can be written:
 \begin{center}
{\small $(\mathrm L_{||}\times\mathrm E_\bot)\times <\bvec> =(\mathrm L_{||}\times <\bvec>)\times\mathrm E_\bot
=(\mathrm L_{||}\times <\bvec_{||}>)\times\mathrm E_\bot=\mathrm G_{||}\times\mathrm E_\bot$,} 
\end{center}
the penultimate equality resulting from $\bvec_{\bot} \in \mathrm E_\bot$. By definition,
G$_{||}=\mathrm L_{||}\times <\bvec_{||}>$ is the usual physical glide plane; thereby our definition of the glide
manifold G is consistent with the definition of the glide plane G$_{||}$ of the
physical dislocation line. Also, $\mathrm G_{||}=\mathrm L _{||}\times\bvec_e$.\smallskip

Consider now the modification of the cut surface  $\Sigma_{||}\subset \mathrm E_{||}$ under glide and the effect on $\Sigma_{\bot}\subset \mathrm E_\bot$, e.g. in Fig.~\ref{fig4}. As L$_{||}$ is displaced along $\bvec_e$, it meets a certain number of atoms $\{m\}$ $-$ added to the cut surface $\Sigma_{||}$ $-$ that carry atomic surfaces of centers $\mvec$; are also added, by interpolation as above, virtual atomic surfaces attached to the continuous positions visited by the moving L$_{||}$. This allows us to define a continuous set of projections $\mvec_\bot$ on AW(p). If the displacement of L$_{||}$ takes a 'full' value $\bvec_e$ (or $\bvec_{||}$, which is equivalent), the displacement in AW(p) is equal to $\bvec_\bot$, i.e. necessarily a false site for $\bm\beta ''= \bm\beta ' + \bvec_\bot$, which result is visible from the inspection of Fig.~\ref{fig4}b; by interpolation there is a continuous path between $\bm\beta '$ and $\bm\beta ''$ which is entirely in $\Fcal$. Thus 'phason' defects (imperfect dislocations) are generically produced by glide.
\subsection{Climb}

Climb is of another nature. Pure climb in the hypercrystal is a displacement of the hyperline L along a well defined direction <{\bf c}> that is
perpendicular both to L and to $\bvec$. Since $\mathrm E_\bot$
belongs to L, <{\bf c}> is perpendicular to $\mathrm E_\bot$, and thus belongs to the physical space $\mathrm E_{||}$. In order to fully achieve the orthogonality of <{\bf c}> to $\bvec$ and to L, it is then enough that <{\bf c}> be perpendicular to $\bvec_{||}$ and to $\mathrm L_{||}$.
Therefore pure climb in physical space is along <{\bf c}>, and is thus the same
process as pure climb in the hypercubic lattice. 

Let $\bm\gamma_{||}$ be the amount of climb along <{\bf c}>; any atom met along <{\bf c}> (real or interpolated) is defined by its position $\bm\gamma_{||}$ and its atomic surface with center $\bm\gamma =\bm\gamma_{||}+\bm\gamma_{\bot}$. By the definition of climb in the hyperlattice, one has $\bm\gamma\cdot\bvec =0$; likewise, by the definition of climb in the physical space, $\bm\gamma_{||}\cdot\bvec_{||}=0$. Thus one 
gets \begin{center}
$\bm\gamma_{\bot}\cdot\bvec_{\bot}=0$;
\end{center}
 thus the displacement of $\bm\beta '$ to $\bm\beta ''$ ($\bm\beta '' = \bm\beta ' + \bm\gamma_{\bot}$) locates $\bm\beta ''$ close to the boundary between $\Tcal$ and $\Fcal$, either in $\Fcal$ or in $\Tcal$. According to the sign of $\bm\gamma_{\bot}$, climb is accompanied by the formation of imperfect dipoles (a small amount) or no dipoles at all. This seems to  indicate that climb is more favored in one direction.

\section{Discussion}\label{discu}  

 An important result of this article is the demonstration of the partition in physical space of a general dissection (a general dislocation $\bvec = \bvec_{||} + \bvec_\bot$ into perfect and  imperfect parts. It is indeed what is observed: the so-called phason components are always separated from the main (perfect) dislocation line, in all the experimental observations as well as in the simulations. In fact, it is the way these results are discussed in the literature, where one refers often to a cloud of phasons accompanying the main dislocation; furthermore in certain experiments a total annealing of this cloud has been observed, just leaving the perfect component alone. It is most surprising that these results haven't yet been qualified as resulting necessarily from some sort of separation of the dislocation line in the hyperspace representation.
 
 Another result is the nature of the stacking fault bordered by an imperfect dislocation, namely its relationship with flipping vectors. This suggests to measure the Burgers vector and the stacking fault translation vector of imperfect dislocations. 
 
 The $\bvec_{||}$ dislocations are singularities {\it per se}, since the cloud of accompanying matching faults can be erased by opposite matching faults or by diffusion \cite{feuerbacher06b}. {Thus there is a topological classification of the dislocations $\bvec_{||}$ alone \cite{kleman92}.} Similarly, as pointed out in (I), there is a topological classification of the matching faults. These two classifications can be given a unique framework for $\bvec$ \cite{kleman92}: but whereas crystal dislocation lines relate to a group of commutative translation symmetries, quasicrystal disvections relate to a group of non-commutative {\sf transvection symmetries} (Cartan's \cite{cartan63}), thus their name. The VP provides a simpler, more physical picture. \smallskip
 
Because the 'phason' defects are
imperfect dislocation dipoles that relax the long range 'phonon' stresses of perfect dislocations, glide in a QC shows some analogy with
Kronberg's synchroshear \cite{kronberg57}, where dislocation glide is assisted by shifts of the smaller atoms, yielding partial companion dipoles. However, since climb is an easier process than glide,
 the question remains open whether the formation of imperfect dipoles is easier by synchroclimb rather than by synchroshear, or whether a high activation energy for synchroclimb definitely makes certain that climb is favored in one direction only.\smallskip

 The octagonal case has often been considered from a conceptual viewpoint, thanks to its (relative) simplicity; this should encourage experiments (observation of defects) in this type of structure.
 
  \section*{Acknowledgments} I thank Jacques Friedel for encouragements $-$ for this part and for part (I).

 This is IPGP contribution \# 3384.
 

  \section*{Appendix A}

\begin{figure}[htbp] 
   \centering
   \includegraphics[width=4in]{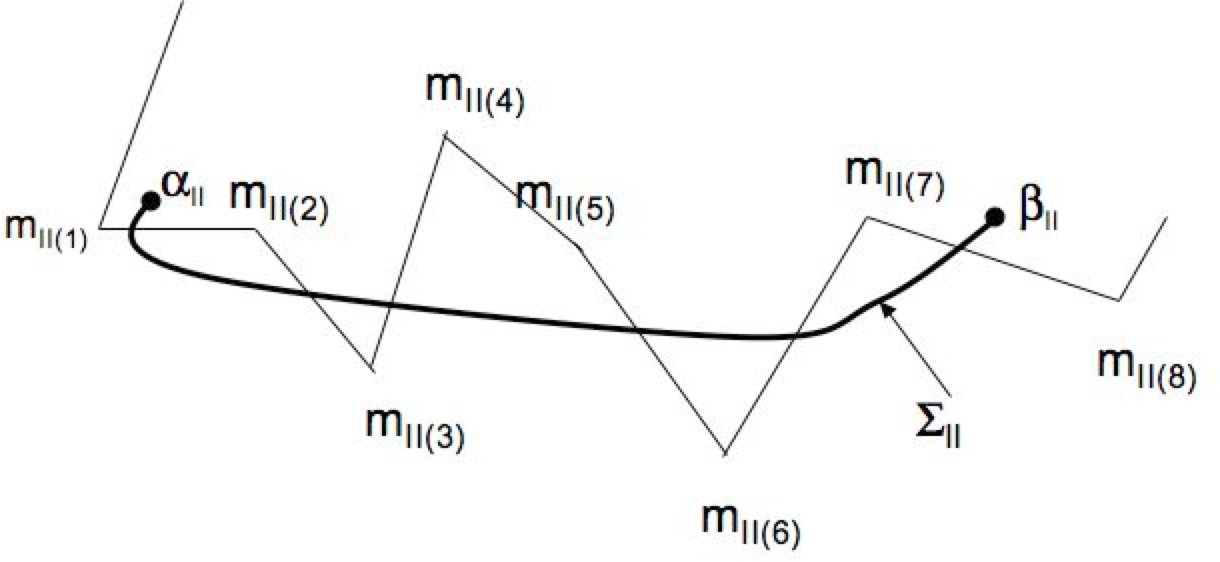} 
   \caption{The polygonal line $\mathrm m_{||(1)}\ \mathrm m_{||(8)}$ is an approximation of the cut surface (in fact a line) $\Sigma_{||} \subset \mathrm E_{||}$ in a 2D quasicrystal, which extends between the dislocation dipole elements $\alpha_{||}\ \beta_{||}$. Its image $\Sigma_{\bot}$, which is generally tortuous, terminates on two points $\alpha,\ \beta$ that are the images in $\mathrm E_\bot$ of $\alpha_{||},\ \beta_{||}$, as in Fig. \ref{fig4}.}
   \label{fig9}
\end{figure}
The demonstration of the continuity of the image $\Sigma_\bot \subset \mathrm E_{\bot}$  of the cut surface $\Sigma_{||}\subset  \mathrm E_{||}$ goes as follows. Consider Fig.~\ref{fig10} which is a representation of the cut surface in $\mathrm E_{||}$ for the 2D case (thus including octagonal and Penrose quasicrystals). The segments m$_{||(i)}$ m$_{||(i+1)}$ are the edges of the quasilattice that cross $\Sigma_{||}$, supplemented by those necessary to construct a continuous (open) polygon which forms an approximation of $\Sigma_{||}$. These are projections of edges m$_{(i)}$ m$_{(i+1)}$ of the hyperlattice in E. The polygonal \textsl{continuous} sequence made of these edges  m$_{(i)}$ m$_{(i+1)}$ necessarily projects orthogonally \textsf{inside} the acceptance window AW $\subset \mathrm E_\bot$ $-$ this is ensured by the construction itself of a quasilattice by the strip and projection method $-$ along a \textsl{continuous} polygonal sequence of segments m$_{\bot (i)}$ m$_{\bot (i+1)}$. Of course this sequence can be much tortuous and display intersections in AW, coming from different points in 
E$_{||}$, but this is a complication which does not invalidate the main results, namely that the cut surface (here a cut segment) can be partitioned into cut subsurfaces (hereinto cut subsegments), respectively in $\Tcal$ and $\Fcal$.

To complete the demonstration, it remains to remark that this sequence of segments m$_{||(i)}$ m$_{||(i+1)}$ can be transformed smoothly into the cut surface $\Sigma_{||}$ $-$ this is the interpolation process of Sect.~\ref{tfp} $-$, and its mapping in E$_{\bot}$ be transformed equally smoothly into a cut surface $\Sigma_{\bot}$. Thus $\Sigma_{||}$ is mapped to a continuous 
$\Sigma_{\bot}$. The extension to a 3D quasicrystal would start from a continuous 2D polyhedral surface constructed on the edges of the quasi lattice that cross $\Sigma_{||}$, then lifted to the hyperlattice (with no breaking of the continuity), eventually projected orthogonally inside the acceptance window, with no breaking either of the continuity.

 \section*{Appendix B}

$\Rcal_1$ and $\Rcal_3$ being equivalent under a transposition, we turn our attention to $\Rcal_1$ only. The stages of the VP, i.e., the displacement of the atoms on the cut surface, are sketched in Fig.~\ref{fig9}. There are in fact two cases:
\begin{figure}[h] %
   \centering
   \includegraphics[width=4.4in]{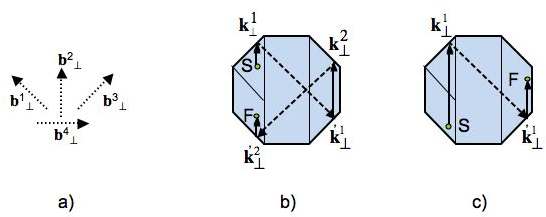} 
   \caption{a) directions of the flipping vectors; b) $\mvec_{\bot\S} in\Rcal_1'$. There are two flips: $-\bvec_{\bot}^1$ and $-\bvec_{\bot}^3$, $\bvec^{eff} =\bvec^2 -\bvec^1-\bvec^3$, $\mathbf{SF} = (1-\sqrt 2)\ \bvec_{\bot}^2$; c) $\mvec_{\bot\S} \in\Rcal_1''$. One flip only: $-\bvec_{\bot}^1$, $\bvec^{eff} =\bvec^2 -\bvec^1$.}
 \label{fig10}
\end{figure} 

\noindent $-$ $\mvec_\bot$ is in S (starting point) in $\Rcal'_1$, the triangular region in the upper part of $\Rcal_1$. In that case the atom meets the boundary of AW after a rather short run, and two flips are necessary for a displacement $\bvec_{\bot }^2=  \bf S\kvec^1_\bot+\kvec^{'1}_\bot \kvec^2_\bot+\kvec^{'2}_\bot \bf F$. The effective Burgers vector, flips included, is 
\begin{equation}
\label{e3}
\bvec _{||}^{\scriptsize eff}=\bvec_{||}^2 -\bvec_{||}^1-\bvec_{||}^3 
\end{equation}in the physical space E$_{||}$. The flips define the matching fault shift,

\noindent $-$ $\mvec_\bot$ is in S (starting point) in $\Rcal''_1$, the parallelogram in the lower part of $\Rcal_1$. Then $\bf S\kvec^1_\bot$ is larger than in the former case, and the displacement $\bvec_{2\bot} = 
\bf S\kvec^1_\bot+\kvec^{'1}_\bot \bf F$. There is only one flip, and the effective Burgers vector, flip included, is
\begin{equation}
\label{e4}
\bvec_{||}^{eff} =\bvec_{||}^2 -\bvec_{||}^1.
\end{equation}

Here again, the matching fault shift is a flip.

\scriptsize

\end{document}